\documentclass[15]{article}
\usepackage{amsmath}
\usepackage{graphics}
\usepackage{graphicx}
\usepackage{epsfig}

\newcommand{\Ket}[1]{\left\vert#1\right\rangle}

\newcommand{\KetBra}[2]{\left\vert#1\right\rangle\left\langle#2\right\vert}
\newcommand{\MatrixEl}[3]{\left\langle#1\right\vert #2 \left\vert#3\right\rangle}
\newcommand{\MeanValue}[1]{\left\langle#1\right\rangle}

\begin{document}
 \parbox{13 cm}
 {
 \begin{flushleft}
 \vspace* {1.2 cm}
 {\Large\bf
 {Revealing non-classical behaviours in the oscillatory motion of a trapped ion
 }
 }\\
 \vskip 1truecm
 {\large\bf
 {
 B.Militello$^1$, A.Messina$^1$, A.Napoli$^1$
 }
 }\\
 \vskip 5truemm
 {
 $^1$)
INFM, MIUR
and Dipartimento di Scienze Fisiche ed Astronomiche \\Via Archirafi 36, 90123 Palermo (ITALY) %\footnote{e-mail: messina@fisica.unipa.it} 
\\e-mail: messina@fisica.unipa.it
 }

 \end{flushleft}
 }
 \vskip 0.5truecm
 {\bf Abstract:\\}
 {
 \noindent
The possibility of revealing non-classical behaviours in the dynamics of a trapped ion via measurements of the mean value of suitable operators is reported. 
In particular we focus on the manifestation known as \lq\lq Parity Effect\rq\rq which may be observed \emph{directly measuring} the expectation value of an appropriate correlation operator.
The experimental feasibility of our proposal is discussed.
 }

%%%%%%%%%%%%%%%%%%%%%%%%%%%%%%%%%%%%%%%%%%%%%%%%%%%%%%%%%%%%%%%%%%%%%%%%%%%%%

 \section{Introduction}

Trapped ions furnish a physical scenario wherein many interesting applications may be realized and quantum mechanical manifestations observed.
In a Paul trap, an ion is confined in such a way that its center of mass is substantially nothing but a quantized tridimensional harmonic oscillator, the three quadratic well frequencies being determined by the ion properties and trapping field parameters\cite{nist,Toschek}.

Subjecting the trapped ion to suitable laser field configurations, it is possible to generate a wide variety of vibronic couplings involving both the center of mass degrees of freedom and the internal atomic state variables\cite{nist,Vogel-Rass,Vogel-Modelli}.
Under the action of such interactions the ion, prepared in appropriate dynamical configurations, exhibits non-classical features. We will focus our attention on a particular quantum mechanical effect known as \lq\lq Parity Effect\rq\rq, brougth to light in cavity QED\cite{Parity-Anna} and then found in trapped ions systems too\cite{Parity-Sabry}.
The detection of this effect may be performed using, for example, methods already proposed for reconstructing the Wigner distribution\cite{Davidovich}. In this way, one con observe all dynamical manifestations of the system and in particular the Parity Effect. Nevertheless this effect also reveals itself in a simpler way, for example in the temporal evolution of the mean value of an appropriate correlation operator relative to two components of the center of mass vibrations. Such a correlation exhibits very different behaviours depending on the parity of the initial number of vibrational quanta. Hence, in some sense, the \lq\lq observation\rq\rq of just the expectation value of a suitable correlation operator is enough to clearly reveal the occurrence of the effect we are interested in.

In this paper we present a method to perform a \emph{direct measurement} of the mean value of a correlation operator appropriate to reveal the Parity Effect. We also show that our procedure is applicable to other vibrational variables.
As direct measurement we mean the possibility of obtaining just the expectation value of an observable without reconstructing the eigenvalue probability distribution at all.

In the next section we will describe in detail the occurrence of the Parity Effect and in the subsequent section we will present our approach to vibrational variables expectation value measurements focusing on the correlation operator appropriate for Parity effect detection. Finally, in the last section, we will give some conclusive remarks and the generalization of the presented procedure.

 \section{Non-Classical behaviours: Parity Effect}

Consider a two level ion confined in a Paul trap, so that its free dynamics is governed by the Hamiltonian
\begin{equation} \label{unperturbed}
  \hat{H}_0=\hbar\sum_{j=x,y,z}\omega_{j}\hat{a}_j^{\dag}\hat{a}_j
           +\hbar\omega_{A} \hat{\sigma}_{z}
\end{equation}
Here $\omega_j$ ($j=x,y,z$) are the frequencies of the trap, $\hat{a}_j(\hat{a}_j^{\dag})$ with $j=x,y,z$ are the annihilation (creation) operators of the centre of mass oscillatory motion. The axial symmetry of the trap implies $\omega_{x}=\omega_{y}$.
Moreover $\omega_A$ is the atomic transition frequency and $\hat{\sigma}_z$ the $z$ component of the pseudospin operator $\stackrel{\rightarrow}{\sigma}\equiv(\hat{\sigma}_x,\hat{\sigma}_y,\hat{\sigma}_z)$ associated to the internal ionic dynamics. Here $\hat{\sigma}_x=\hat{\sigma}_+ +\hat{\sigma}_-$, $\hat{\sigma}_y=i(\hat{\sigma}_+ -\hat{\sigma}_-)$, $\hat{\sigma}_+=\KetBra{+}{-}$,
$\hat{\sigma}_-=\KetBra{-}{+}$, $\hat{\sigma}_z=\KetBra{+}{+}-\KetBra{-}{-}$,
and $\Ket{\pm}$ are the two effective atomic levels.

Prepare the ion in a vibrational $SU(2)$ coherent state
\begin{equation} \label{SU2-state}
  \Ket{\psi_{in}}=\sum_{k=0}^{N}\frac{1}{2^{\frac{N}{2}}}\left(\frac{N!}{(N-k)!k!}\right)^{\frac{1}{2}}\Ket{n_x=k, n_y=N-k, n_z=0}\Ket{-}
\end{equation}
This is a state with a well defined bosonic excitation numberm, $N$, in the $xy$ plane, which is nothing but a Fock state along the bisector of the first $xy$-quadrant.
Subject now the particle to a suitable laser configuration responsible for the \emph{two-mode and two-phonon Jaynes-Cummings model}\cite{Parity-Sabry}
\begin{equation} \label{JC-Model}
  \hat{H}_{int}=\hbar g \left(\hat{a}_x\hat{a}_y\hat{\sigma}_+ + h.c.\right)
\end{equation}

The dynamics of the trapped ion under such conditions has been studied in detail in ref\cite{Parity-Sabry} where has been brought to light that, at certain instant of time, the system discriminate between the two cases corresponding to \lq\lq $N$ even\rq\rq or \lq\lq $N$ odd\rq\rq. Such a discrimination reveals itself, for example, in the behaviour of the mean value of an operator which correlates the two motions along $x$ and $y$, that is
\begin{equation} \label{C_xy}
  \hat{C}_{xy}=\hat{a}_x^{\dag}\hat{a}_y+\hat{a}_x\hat{a}_y^{\dag}
\end{equation}
or its square, $\hat{C}_{xy}^2$.

The mean value of $\hat{C}_{xy}$ is evaluated in the state of the ion after an electronic \emph{conditional measurement} has been performed. More precisely, we consider the unitary time evolution of the system under the action of the hamiltonian (\ref{JC-Model}) and then induce a collapse of the electronic state (observing it). If the ion is found in the atomic ground state, we consider the mean value $\MeanValue{\hat{C}_{xy}}$ in the collapsed wave function. 
In fig.1 are shown the graphics of $\MeanValue{\hat{C}_{xy}}$ evaluated at time $t$ for two different values of $N$: $N=20$ and $N=21$.
It is well visible that there exists an instant of time at which the mean value of the correlation operator $\hat{C}_{xy}$ assume the value $(-1)^N N$, discriminating in a \emph{mesoscopic} way (a difference almost equal to $2N=40$ in the mean value) between two just \emph{microscopically} distinguishable initial conditions (different just for $1$ vibrational quantum).

% Figure 
      \begin{figure}
        \begin{center}
          \includegraphics[width=4cm, angle=-90]{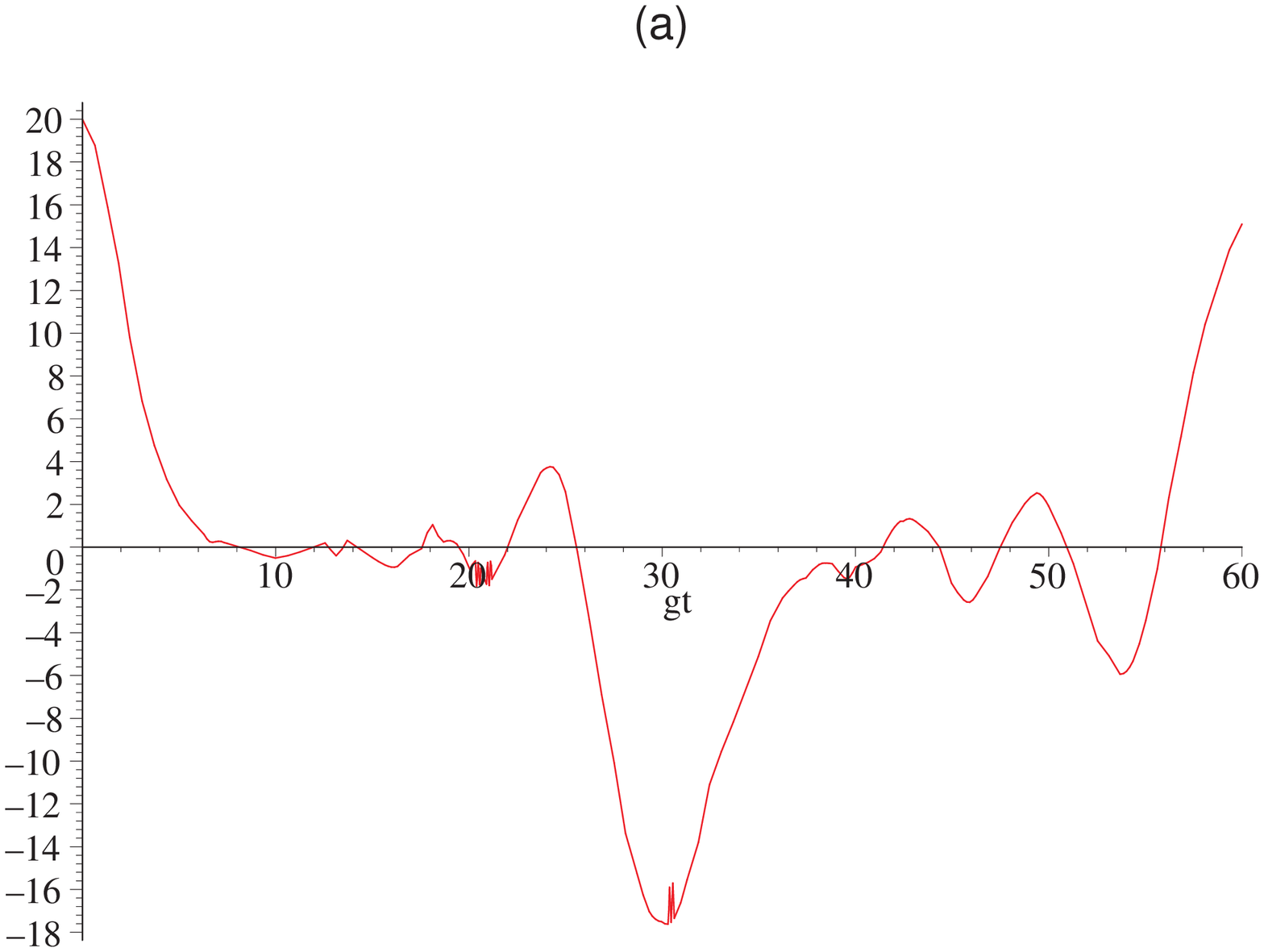}
        \end{center}
        %\bigskip
        \begin{center}
          \includegraphics[width=4cm, angle=-90]{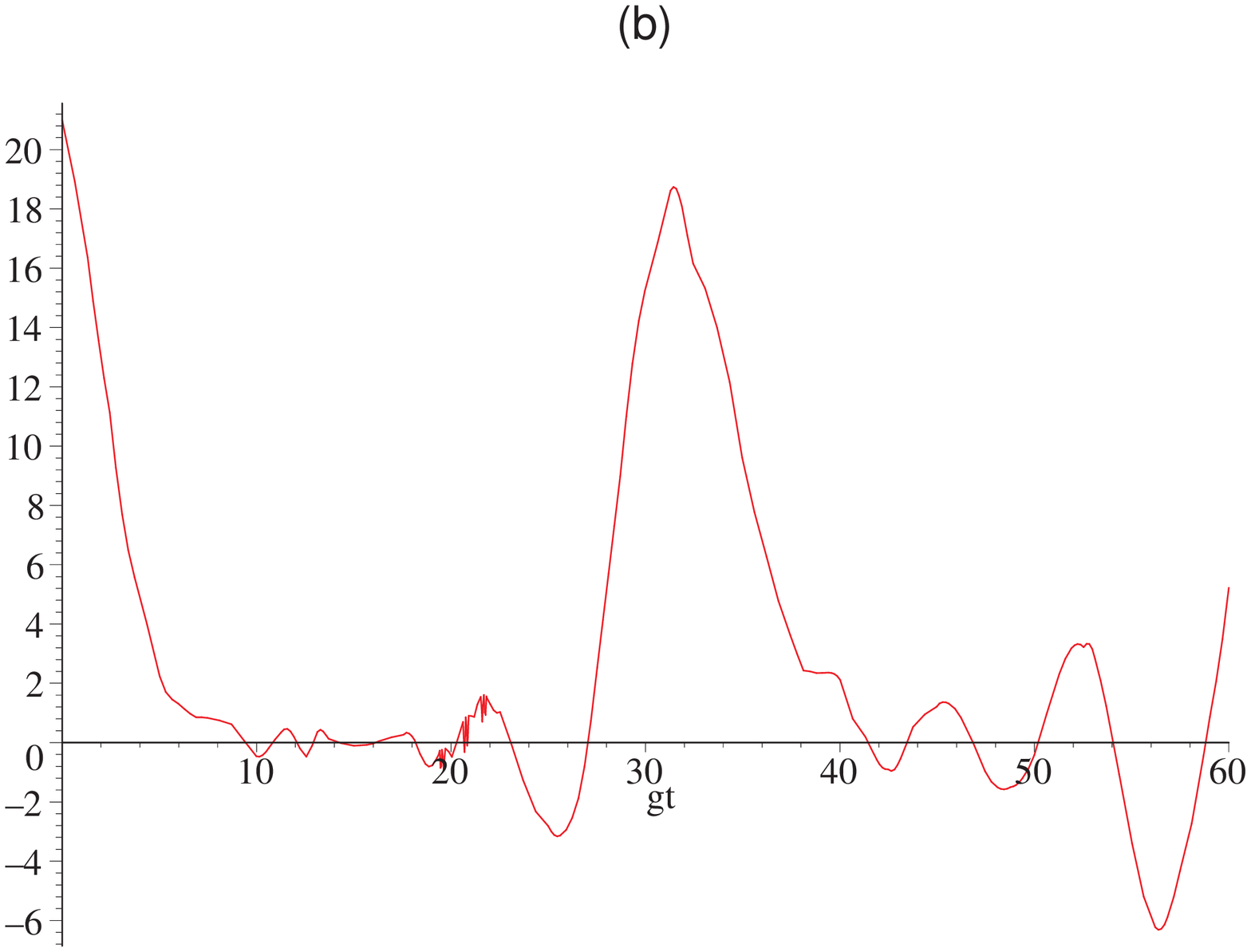}
        \end{center}
        \caption{\footnotesize   
Temporal evolution of $\langle\hat{C}_{xy}\rangle$ for $N$=20 (a) e $N$=21 (b).}
      \end{figure}

 \section{Direct measurement of $\MeanValue{\hat{C}_{xy}}$}

In order to directly measure the mean value of the correlation operator $\hat{C}_{xy}$ one can use the procedure outlined in ref\cite{My-Work}. This requires the implementation of a suitable vibronic coupling 
\begin{equation} \label{interaction}
  \hat{H}^{(I)}_I=\hbar\gamma\hat{C}_{xy}\hat{\sigma}_{x}
\end{equation}
which induces atomic transitions whose speed is determined by the mean value of the motional operator $\hat{C}_{xy}$. Hence, monitoring the electronic state after the action of hamiltonian (\ref{interaction}), it is possible to obtain information about the mean value of $\hat{C}_{xy}$.

The vibronic coupling (\ref{interaction}) may be realized just considering two lasers directed along the axis $x'$ and $y'$ which are $x$ and $y$ axis $\frac{\pi}{4}$-rotated about $z$ respectively. These two lasers must have the same intensities, be $\pi$ out of phases and both tuned to the electronic transition frequency, $\omega_A$.
The evaluation of the effective hamiltonian in the Lamb-Dicke limit and in the Rotating Wave Approximation just leads to the hamiltonian in eq.(\ref{interaction}).

Let us consider now the dynamics induced by this interaction.
If the ion is described by the state
\begin{equation} \label{init-cond-z}
  \Ket{\psi(0)}=\Ket{\psi_{vibr}}\Ket{-}
\end{equation}
one may \emph{rotate} the electronic state (via a \emph{non dispersive} interaction due to a Raman laser tuned to $\omega_A$, that is via a hamiltonian of the form $\hat{H}_{I}\propto \delta \hat{\sigma}_x$) in order to obtain
\begin{equation} \label{init-cond-y}
  \Ket{\overline{\psi}(0)}=\Ket{\psi_{vibr}}\Ket{-}_y
\end{equation}
where $\Ket{-}_y$ is an eigenstate of the operator $\hat{\sigma}_y$. Applying now the interaction in eq.(\ref{interaction}) and evaluating the mean value of $\hat{\sigma}_z$ one obtains
\begin{eqnarray}\label{Mean-Sigma-L}
  %\nonumber
  \MeanValue{\hat{\sigma}_z(t)}
  %&=&
  =
    \MatrixEl{\overline{\psi}(0)}
             {e^{i\frac{\hat{H}^{(I)}_I}{\hbar}t}
              \hat{\sigma}_z
              e^{-i\frac{\hat{H}^{(I)}_I}{\hbar}t}}
             {\overline{\psi}(0)}=%\\
  %&=&
    -\MatrixEl{\psi_{vibr}}{\sin(2\gamma t \hat{C}_{xy})}{\psi_{vibr}}
\end{eqnarray}

Suppose now that the mean value of the operator $\sin(2\gamma t \hat{C}_{xy})$ may be linearized, meaning that the following approximation is valid
\begin{equation} \label{linearization}
  \MeanValue{\sin(2\gamma t \hat{C}_{xy})} \simeq \MeanValue{2\gamma t \hat{C}_{xy}}
\end{equation}

Under such a hypothesis the mean value of $\hat{\sigma}_z$, evaluated after the action of the hamiltonian in eq.(\ref{interaction}), is proportional to the mean value of $\hat{C}_{xy}$ before such a vibronic interaction.

Observe now that in typical experiments performed with trapped ions, the number of bosonic excitations doesn't exceed 30. Moreover, since the correlation operator $\hat{C}_{xy}$ is canonically equivalent to the operator $\hat{a}_y^{\dag}\hat{a}_y-\hat{a}_x^{\dag}\hat{a}_x$ (as well visible transforming it considering a $\frac{\pi}{4}$-rotation about z), we can assume that no eigenstate of $\hat{C}_{xy}$ corresponding to an eigenvalue $c>c_{max}=30$ is involved in the series expansion of $\Ket{\psi_{vibr}}$. 
This is a central assumption of our protocol that legitimates the linearization of the sinusoidal operator.
Indeed, in this situation we can choose the interaction time relative to the hamiltonian in eq.(\ref{interaction}) in such a way that $\sin(2\gamma t c)\simeq 2\gamma t c$ for $|c|<c_{max}$ and this easily leads to the approximation in eq.(\ref{linearization}).

The last step of our analysis consists in estimating the necessary duration, $t$, of the interaction due to hamiltonian (\ref{interaction}), and in verifying its consistency with decoherence and typical experimental times. 
To this end consider now characteristic values for the coupling strength, for instance $\gamma\sim 10 KHz$\cite{exp-values-nist,exp-values-Inn}, and $0.4$ as the bound of sinus linearizability zone (meaning that we assume $\sin x\simeq x$ for $|x|<0.4$). From these assumptions we deduce an interaction time of the order of magnitude of $1\mu s$, which is compatible both with decoherence (in the sense that coherences are maintained during the vibronic interaction) \cite{exp-values-nist}
and \lq\lq laser switch-on/switch-off\rq\rq typical times.

 \section{Conclusive remarks}

Summarizing we have recalled the Parity Effect occurring in the dynamics of a trapped ion under some particular conditions, stressing the fact that such a behaviour may be revealed considering the mean value of a suitable correlation operator $\hat{C}_{xy}$.
Then we have presented a proposal for measuring the mean value of a suitable correlation operator \emph{without reconstructing its eigenvalue probability distribution function at all}. 

The possibility of implementing the vibronic coupling (\ref{interaction}) and the values of the quantities involved in the protocol above stated (strength coupling, interaction time, maximum value of the eigenvalues of $\hat{C}_{xy}$) both imply a good degree of experimental feasibility.

We conclude this paper stressing that the mean value measurement protocol here presented is also applicable to a wide variety of motional observable like for example, energy, position, momentum, angular momentum\cite{My-Work}. Indeed, our procedure substantially rests upon the implementation of the hamiltonian (\ref{interaction}) \emph{mutatis mutandis} and in the individuation of a \emph{cut-off} for the eigenvalues of the observable under scrutiny. This last assumption is valid for vibrational energy (we have already recalled the maximum value of 30 excitations), obviously for position (by definition of trapped ion) and even for angular momentum (which, as the correlation operator, may be unitarily transformed into the difference of two phononic number operators).

 \section{Acknowledgements}
We thank S.Maniscalco for stimulating discussions.
One of the authors (A. N.) acknowledges financial support from Finanziamento Progetto Giovani Ricercatori anno 1998, Comitato 02.

 %References
 
 \end{document}